\documentclass[12pt]{article}
\usepackage{amsmath}
\usepackage{graphicx,psfrag,epsf}
\usepackage{enumerate}
\usepackage{url} 

\newcommand{\blind}{0}

\usepackage{amsthm,amsfonts}
\usepackage{comment}
\usepackage[bf]{caption}
\usepackage{lscape}
\usepackage[backend=bibtex, style=nature, citestyle=authoryear]{biblatex}
\bibliography{bibliography}

\def\inprob{\stackrel{p}{\rightarrow}}
\def\indist{\rightsquigarrow}

\def\T{{ \mathrm{\scriptscriptstyle T} }}

\newcommand{\Pb}{P}
\newcommand{\Pn}{\mathbb{P}_n}
\newcommand{\eff}{\text{\scriptsize{eff}}}
\newcommand{\E}{\mathbb{E}}
\newcommand{\R}{\mathbb{R}}

\DeclareSymbolFont{bbold}{U}{bbold}{m}{n}
\DeclareSymbolFontAlphabet{\mathbbold}{bbold}

\theoremstyle{definition}

\theoremstyle{remark}

\begin{document}

\def\spacingset#1{\renewcommand{\baselinestretch}%
{#1}\small\normalsize} \spacingset{1}


\if0\blind
{
  \title{ \bf Semiparametric theory}
  \author{\\ Edward H. Kennedy\thanks{Edward Kennedy (edward@stat.cmu.edu) is Assistant Professor in the Department of Statistics, Carnegie Mellon University, Pittsburgh, PA 15213. } \\
    Carnegie Mellon University, PA, USA \\ \\ 
    }
  \maketitle
  \setcounter{page}{0}
  \thispagestyle{empty}
} \fi

\if1\blind
{
  \vspace*{.8in}
  \begin{center}
    {\LARGE\bf Semiparametric theory}
\end{center}
  \setcounter{page}{0}
  \medskip
} \fi

\begin{abstract}
In this paper we give a brief review of semiparametric theory, using as a running example the common problem of estimating an average causal effect. Semiparametric models allow at least part of the data-generating process to be unspecified and unrestricted, and can often yield robust estimators that nonetheless behave similarly to those based on parametric likelihood assumptions, e.g., fast rates of convergence to normal limiting distributions. We discuss the basics of semiparametric theory, focusing on influence functions.
\end{abstract}

\noindent%
\vfill

\thispagestyle{empty}

\newpage

\spacingset{1}



In this paper we give a review of semiparametric theory, using as a running example the common problem of estimating an average causal effect. Our review draws heavily on foundational work in general semiparametrics by \textcite{begun1983information, bickel1993efficient, pfanzagl1982contributions, van2000asymptotic,van2002semiparametric}, among others \autocite{newey1994asymptotic, kosorok2007introduction}, as well as many modern developments in missing data and causal inference problems by Robins \& van der Laan \autocite{robins1986new, robins1994estimation, robins1995analysis,  robins2000marginal, van2003unified, van2011targeted, van2014higher, robins2017minimax}, and colleagues \autocite{hahn1998role, tsiatis2006semiparametric}. We refer to \textcite{tsiatis2006semiparametric} for a very readable review with more details.

\section{Setup}

A standard setup in semiparametric theory is as follows. We suppose we observe a sample of independent and identically distributed observations $(Z_1,...,Z_n)$ distributed according to some unknown probability distribution $\Pb$. Then our goal is estimation and inference for a real-valued target parameter, or functional, $\psi=\psi(\Pb) \in \R^q$. 

In this paper we focus on an example where we observe $n$ observations of $Z=(X,A,Y)$, with $X \in \R^p$ covariates, $A$ a binary treatment (or missingness) indicator, and $Y$ an outcome of interest, and our goal is to estimate the ``treatment effect''
\begin{equation}
\psi(\Pb)=\E\{\E(Y \mid X,A=1)\} = \int \E(Y \mid X=x, A=1) \ d\Pb(x)   \label{eq:psi}
\end{equation}
with $\E=\E_\Pb$.  Under causal assumptions such as no unmeasured confounding, the statistical parameter $\psi$ represents the causal quantity $\E(Y^1)$, i.e., the mean outcome had everyone taken treatment. If we let $Y=AY^*$ then $\psi$ also represents the mean of the partially observed outcome $\E(Y^*)$ under a missing at random assumption, where here $A$ is a missingness indicator. In what follows we write $p(V=t)$ for the density of a variable $V$ at $t$, but when there is no ambiguity we let $p(v) = p(V=v)$.

Other archetypal functionals considered in the literature include:
\begin{itemize} \setlength\itemsep{0em}
\item integral density functionals: $\E[h\{p(Z)\}]$, e.g., $h(t)=t$ gives the integrated square density and $h(t)=-\log(t)$ gives the entropy
\item instrumental variable effect: $\frac{\E\{ \E(Y \mid X,V=1)-\E(Y \mid X,V=0)\}}{\E\{ \E(A \mid X,V=1)-\E(A \mid X,V=0)\}}$
\item hazard ratio: $\frac{\lambda(t \mid X=x)}{\lambda(t \mid X=0)}$ where $\lambda(t \mid X) = \lim_{\delta \rightarrow 0} \Pb(t \leq T \leq t+\delta \mid T \geq t,X)/ \delta $
\item optimal treatment regime value: $\max_{d} \E\{ \E(Y \mid X,A=d(X))\}$ over all $d : \mathcal{X} \mapsto \{0,1\}$
\item g-formula: $\E[\E\{ \E(Y \mid X_2,A_2=a_2,X_1,A_1=a_1) \mid X_1,A_1=a_1\}]$
\end{itemize}
The central feature of semiparametrics is that at least part of the data-generating process $\Pb$ can be unrestricted or unspecified. This is crucial because knowledge of the true distribution $\Pb$ is typically lacking in practice, especially when the observations $Z$ include numerous and/or continuous components. Luckily, it turns out that in many functional estimation problems there exist estimators that are $\sqrt{n}$-consistent and asymptotically normal, even in large nonparametric models that put minimal restrictions on $\Pb$. In other words, estimating functionals $\psi(\Pb)$ is typically an easier statistical problem than estimating all of $\Pb$. 

\section{Semiparametric Models}

A \textit{statistical model} $\mathcal{P}$ is a set of possible probability distributions, which is assumed to contain the true observed data distribution $\Pb$. Using a parametric model amounts to assuming the true distribution is known up to a finite-dimensional real-valued parameter $\theta \in \R^q$, e.g., we may have $\mathcal{P}=\{ \Pb_\theta : \theta \in \R^q\}$ with $\psi \subseteq \theta$. For example, if $Z$ is a scalar random variable one might assume it is normally distributed with unknown mean and variance, $Z \sim N(\mu,\sigma^2)$, in which case the model is indexed by $\theta=(\mu,\sigma^2) \in \R \times \R^+$. \textit{Semiparametric models} are simply sets of probability distributions that cannot be indexed by only a Euclidean parameter, i.e., models that are indexed by an infinite-dimensional parameter in some way. Semiparametric models can vary widely in the amount of structure they impose; for example, they can range from \textit{nonparametric models} for which $\mathcal{P}$ consists of all possible probability distributions, to simple regression models that characterize the regression function parametrically but leave the residual error distribution unspecified. 

For the treatment effect functional $\psi=\E\{\E(Y \mid X,A=1)\}$, and in other general causal inference and missing data problems, one may often want to incorporate some knowledge about the treatment mechanism $p(a \mid x)$ but leave the other components $p(y \mid x,a)$ and $p(x)$ unspecified. This is because the covariate/outcome mechanisms are often complex natural processes outside of human control, whereas the treatment mechanism is known in randomized trials, and can be well-understood in some observational settings (for example, when a medical treatment is assigned in a standardized way, which is communicated by physicians to researchers). In an experiment where $p(a \mid x)$ is set to be $0.5$, for example, this amounts to the restriction
$$ \mathcal{P} = \{ \Pb: p(z) = p(y \mid x, a) p(x) / 2 \}  $$
where $\{ p(y \mid x, a), p(x)\}$ are viewed as unspecified infinite-dimensional nuisance parameters.

Of course it is not always the case that there is substantive information available about some component of $P$, such as the treatment mechanism in the above model. In many  studies no parts of the data-generating process are under human control, and all components may be unknown and possibly very complex (e.g., in studies where even the exposure itself is a disease or other medical condition). It would then often be more appropriate to consider inference under a nonparametric model that makes no parametric assumptions about the distribution $\Pb$. For instance, in the treatment effect example, one would thus also allow $p(a \mid x)$ to be an unrestricted nuisance function. However, in order to obtain estimators that converge at $\sqrt{n}$ rates in nonparametric models, nuisance functions will often have to satisfy some structural conditions, such as H\"older smoothness or bounded variation. 

Semiparametric models can also arise via parametric assumptions about non-Euclidean functionals. For example, the causal assumptions that identify $\psi$ also imply $\E(Y^1 \mid V) = \E\{ \E(Y \mid X,A=1) \mid V\}$ for any $V \subseteq X$; thus one might employ a parametric assumption of the form $\E\{ \E(Y \mid X,A=1) \mid V=v\}=m(v;\beta)$ but leave the rest of $\Pb$ unrestricted. Similarly, the famous Cox proportional hazards model assumes that the hazard ratio follows the parametric form $\frac{\lambda(t \mid X=x)}{\lambda(t \mid X=0)} = \exp(\beta^\T x)$. These restrictions are somewhat similar in spirit to classical parametric models. Unlike the experiment represented by model \eqref{eq:piknown}, the assumptions are not guaranteed by the study design.

\section{Influence Functions}

Here we discuss \textit{influence functions}, foundational objects in nonparametric efficiency theory that allow us to characterize a wide range of possible estimators and their efficiency. There are two notions of an influence function: one corresponds to estimators and one corresponds to parameters. To distinguish these cases we will call the former \textit{influence functions} and the latter \textit{influence curves}; we focus on the former in this section. 

Let $\Pn=n^{-1} \sum_i \delta_{Z_i}$ denote the empirical distribution of the data, with $\delta_z$ the Dirac measure, so that sample averages can be written as $n^{-1} \sum_i f(Z_i) = \int f(z) \ \!d\Pn =\Pn\{ f(Z)\}$. Then an estimator $\hat\psi$ is asymptotically linear and has \textit{influence function} $\varphi$ if it can be approximated by an empirical average of $\varphi$, i.e.,
\begin{equation}
\hat\psi - \psi = \Pn\{ \varphi(Z) \} + o_\Pb(1/\sqrt{n})  \label{eq:inf_fn}
\end{equation}
where $\varphi$ has mean zero and finite variance (i.e., $\E\{\varphi(Z)\}=0$ and $\E\{\varphi(Z)^{\otimes 2}\}< \infty$). Here $o_\Pb(1/\sqrt{n})$ employs the usual stochastic order notation so that $X_n=o_\Pb(1/r_n)$ means $r_n X_n \inprob 0$ where $\inprob$ denotes convergence in probability. 

Importantly, by the central limit theorem, \eqref{eq:inf_fn} implies $\hat\psi$ is asymptotically normal with
\begin{equation}
\sqrt{n}(\hat\psi - \psi_0) \indist N\Big(0, \ \E\{\varphi(Z)^{\otimes 2}\} \Big) ,
\end{equation}
where $\indist$ denotes convergence in distribution. Thus if we know the influence function for an estimator, we know its asymptotic distribution and can easily construct confidence intervals and hypothesis tests. Also, the influence function for an asymptotically linear estimator is almost surely unique, so in this sense the influence function contains all information about an estimator's asymptotic behavior (up to $o_\Pb(1/\sqrt{n})$ error). 

Consider our running example where $\psi$ is defined as in \eqref{eq:psi}. When the propensity score is known to be $p(A=1 \mid X=x) = \pi(x)$, a simple weighting estimator is given by
$$ \hat\psi_{ipw}^* = \Pn\left\{ {AY}/{\pi(X)} \right\} . $$
It is straightforward to check using iterated expectation that $\E(\hat\psi_{ipw}^*) =\psi$. Then the influence function for $\hat\psi_{ipw}^*$ is simply given by
$ \varphi_{ipw}(Z) = AY/\pi(X) - \psi $
since $\hat\psi_{ipw}^* - \psi = \Pn\{\varphi_{ipw}(Z)\}$ exactly, without any $o_\Pb(1/\sqrt{n})$ approximation error. Interestingly, it can be shown  that the estimator $\hat\psi_{ipw}$ that uses an estimated propensity score $\hat\pi$ in place of the true $\pi$ is at least as efficient as the estimator $\hat\psi_{ipw}^*$ that uses the true $\pi$. This follows from the fact that the influence function for $\hat\psi_{ipw}$ equals that of $\hat\psi_{ipw}^*$ minus its projection, so that the variance of the former influence function must be less than or equal to that of the latter, by the Pythagorean theorem. 

Now consider the so-called doubly robust estimator $\hat\psi_{dr} = \Pn\{ \varphi^*_{dr}(Z;\hat\eta) \}$ where
$$ \varphi^*_{dr}(Z;\eta) = \varphi^*_{dr}(Z;\pi,\mu)= \frac{A\{Y-\mu(X)\}}{\pi(X)} + \mu(X) $$
for $\hat\mu(x)$ an estimator of the regression $\mu(x)=\E(Y \mid X=x,A=1)$, and $\eta=(\pi,\mu)$. What is the influence function for $\hat\psi_{dr}$? Consider the decomposition
$$ \hat\psi_{dr} - \psi = (\Pn-\Pb)\Big\{  \varphi^*_{dr}(Z;\hat\eta) -  \varphi^*_{dr}(Z;\eta) \Big\} + (\Pn - \Pb) \varphi^*_{dr}(Z;\eta) + \Pb\Big\{ \varphi^*_{dr}(Z;\hat\eta) -  \varphi^*_{dr}(Z;\eta) \Big\} $$
where  $\Pb\{ \hat{f}(Z) \} = \int \hat{f}(z) \ d\Pb(z) = \E\{ \hat{f}(Z_{n+1}) \mid Z_1,...,Z_n\}$ denotes the expected value of an estimated $\hat{f}$ over a new observation, conditional on the data used to construct it. The above decomposition in fact holds even replacing $\eta$ with  $\overline\eta \in \{ (\overline\pi,\mu), (\pi,\overline\mu), (\pi,\mu)\}$, i.e., if either $\hat\pi$ or $\hat\mu$ is consistent for its true target (not necessarily both), since then $\E\{ \varphi_{dr}^*(Z;\overline\eta)\} = \psi$. This is the famous property called double robustness. 

The first term in the decomposition will be $o_\Pb(1/\sqrt{n})$ under empirical process conditions, e.g., if the estimators $\hat\eta$ are regular enough so that $\varphi_{dr}^*(Z;\hat\eta)$ lies in a Donsker class, or if sample splitting is used so that $\hat\eta$ is constructed on separate data. The second term is simply a centered sample average, and thus converges to a normal distribution after $\sqrt{n}$-scaling, by the central limit theorem. The third term is the really interesting one. For special estimators like $\hat\psi_{dr}$, it can be $o_\Pb(1/\sqrt{n})$ even when nuisance estimators converge at slower nonparametric rates. For example, with $\hat\psi_{dr}$ a sufficient condition is that $\hat\pi$ and $\hat\mu$ converge to $(\pi,\mu)$ at a faster than $n^{1/4}$ rate in $L_2(\Pb)$ norm. Under these kinds of conditions ensuring that the first and third terms in the decomposition are $o_\Pb(1/\sqrt{n})$, the influence function of the estimator $\hat\psi_{dr}$ will be given by $\varphi^*_{dr}-\psi$, since $\hat\psi_{dr} - \psi = \Pn( \varphi^*_{dr}-\psi ) + o_\Pb(1/\sqrt{n})$.

So far we have seen that, given an estimator $\hat\psi$, we can learn about its asymptotic properties by considering its influence function $\varphi(Z)$. But we can also use influence functions to find or construct estimators, for example by solving estimating equations that use the putative influence function as an estimating function. There is a deep connection between (asymptotically linear) estimators for a given model and functional, and the corresponding influence functions. In some sense, if we know one then we know the other. This leads to the notion of an influence function for a parameter $\psi$, which we call an \textit{influence curve}. 

\section{Tangent Spaces \& Influence Curves}

Here we use the term influence curves to denote influence functions for parameters. These are essentially putative influence functions: functions that could be the influence function of a properly constructed estimator, but which may not correspond to an estimator at all, and yet still exist and can be characterized based on the form of the functional $\psi(P)$. First, though, we need to understand tangent spaces and parametric submodels. 

As discussed in the previous section, influence functions $\varphi$ (now called {influence curves})  are functions of the observed data $Z$, and have mean zero and finite variance. Such functions reside in the Hilbert space $L_2(P)$ of measurable functions $g: \mathcal{Z} \rightarrow \R$ with $P g^2 = \int g^2 \ dP = \E\{g(Z)^2\} < \infty$, equipped with covariance inner product $\langle g_1,g_2 \rangle = P(g_1 g_2)$. The space of influence curves will be a subspace of this Hilbert space. A Hilbert space is a complete inner product space, and generalizes usual Euclidean space; it provides a notion of distance and direction for spaces whose elements are potentially infinite-dimensional functions. 

A fundamentally important subspace of $L_2(P)$ in semiparametric problems is the \textit{tangent space}. For parametric models indexed by real-valued parameter $\theta \in \R^{q+1}$, the tangent space $\mathcal{T}$ is defined as the linear subspace of $L_2(P)$ spanned by the score vector, i.e.,
\begin{equation*}
\mathcal{T} = \{ b^\T s_\theta(Z;\theta_0) : b \in \R^{q+1} \}, 
\end{equation*}
where $s_\theta(Z;\theta_0) = \partial \log p(z;\theta) / \partial \theta |_{\theta=\theta_0}$. If we can decompose $\theta=(\psi,\eta)$ then we can equivalently write $\mathcal{T} = \mathcal{T}_\psi \oplus \mathcal{T}_\eta$ for
\begin{equation*}
\mathcal{T}_\psi = \{ b_1 s_\psi(Z;\theta_0) : b_1 \in \R \} \ , \ \mathcal{T}_\eta = \{ b_2^\T s_\eta(Z;\theta_0) : b_2 \in \R^q \}, 
\end{equation*}
where $s_\psi(Z;\theta_0) = \partial \log p(z;\theta) / \partial \psi |_{\theta=\theta_0}$ is the score function for the target parameter, and similarly $s_\eta(Z;\theta_0) = \partial \log p(z;\theta) / \partial \eta |_{\theta=\theta_0}$ is the score for the nuisance parameter ($A \oplus B$ denotes the direct sum $A \oplus B=\{a + b : a \in A, b \in B\}$). In the above formulation, the space $\mathcal{T}_\eta$ is called the \textit{nuisance tangent space}. Influence curves for $\psi$ reside in the \textit{orthogonal complement of the nuisance tangent space}, denoted by $\mathcal{T}_\eta^\perp = \{ g \in L_2(P) : P(g h) = 0 \ \text{ for any } \ h \in \mathcal{T}_\eta \}$. In such parametric settings, this orthogonal space $\mathcal{T}_\eta^\perp$ is
\begin{align*}
\mathcal{T}_\eta^\perp &= \{ g \in L_2(P) : g= h - \Pi(h \mid \mathcal{T}_\eta) , \ h \in L_2(P) \}  \\
&= \{ g \in L_2(P) : g= h - P(h s_\eta^\T) P(s_\eta s_\eta^\T)^{-1} s_\eta ,\  h \in L_2(P) \} , \nonumber
\end{align*}
where $\Pi(g \mid S)$ denotes projections of $g$ on the space $S$, i.e., $P[h \{g-\Pi(g \mid S)\}]=0$ for all $h \in S$. The subspace of influence curves is the set of elements $\varphi \in \mathcal{T}_\eta^\perp$ that satisfy $P(\varphi s_\psi)=1$. The \textit{efficient influence curve} is the influence curve with the smallest covariance $P(\varphi^2)$, and is given by $\varphi_\eff=P(s_\eff^2)^{-1} s_\eff$, where $s_\eff$ is the \textit{efficient score}, given by $s_\eff=s_\psi-\Pi(s_\psi \mid \mathcal{T}_\eta)$.

Thus if we can characterize the nuisance tangent space and its orthogonal complement, then we can characterize influence curves. In fact, one can show that all regular asymptotically linear estimators have influence functions $\varphi$ that reside in $\mathcal{T}_\eta^\perp$ with $P(\varphi s_\psi)=1$, and conversely any element in this space corresponds to the influence function for some regular asymptotically linear estimator. Thus characterizing the nuisance tangent space allows us to also characterize all potential (regular asymptotically linear) estimators. 

In parametric models the tangent space is defined as the span of the score vector $s_\theta$. However, in semiparametric models the nuisance parameter is infinite-dimensional, and so we cannot define scores analogously, as it would require differentiation with respect to this nuisance parameter. How do we extend tangent spaces to infinite-dimensional semiparametric models? The answer lies in a clever device called a parametric submodel.

A \textit{parametric submodel} $\{\Pb_\epsilon : \epsilon \in \R\}$ is a set of distributions contained in a larger model $\mathcal{P}$, which also contains the truth, i.e., $\Pb \in \{ \Pb_\epsilon : \epsilon \in \R\}$. A typical example of a parametric submodel is given by
$$ p_\epsilon(z) = p(z) \{ 1 + \epsilon s(z) \} $$
where $\E\{ s(Z)\}=0$ and $\sup_z |s(z) | \leq M < \infty$ so that $p_e(z) \geq 0$ when $| \epsilon | \leq 1/M$. Note that $s(z)$ is the score function $\frac{d}{d\epsilon} \log p_\epsilon(z) \mid_{\epsilon=0}$ for the above submodel. One intuition behind parametric submodels comes via efficiency bounds. First note that it is an easier problem to estimate $\psi$ under the smaller parametric submodel $\mathcal{P}_\epsilon \in \mathcal{P}$ than it is to estimate $\psi$ under the entire larger semiparametric model $\mathcal{P}$. Therefore the efficiency bound under the larger model $\mathcal{P}$ must be larger than the efficiency bound under any parametric submodel. In fact the efficiency bound for semiparametric models is typically defined in exactly this way, as the supremum of all such parametric submodel efficiency bounds.

Now, just as the tangent space is defined as the linear span of the score vector in parametric models, in semiparametric models the tangent space $\mathcal{T}$ is defined as the (closure of the) linear span of scores of the parametric submodels, i.e., $\mathcal{T} = \{ b^\T s(Z) : b \in \R \}$. Similarly, the nuisance tangent space $\mathcal{T}_\eta$ for a semiparametric model is the set of scores in $\mathcal{T}$ that do not vary the target parameter $\psi$, i.e.,
$$ \mathcal{T}_\eta = \{ g \in \mathcal{T} : \partial \psi(P_{\epsilon})/\partial \epsilon|_{\epsilon=0} = 0 \}. $$
Importantly, in nonparametric models the tangent space is the whole Hilbert space of mean zero functions. For more restrictive semiparametric models the tangent space will be a proper subspace. 

Now we can define influence curves, in much the same way as in parametric models. A parameter $\psi=\psi(P)$ is pathwise differentiable with \textit{influence curve} $\varphi$ if
\begin{equation}
\frac{d}{d\epsilon} \psi(\Pb_\epsilon) \Bigm|_{\epsilon=0} = \int \varphi(z) \left( \frac{d}{d\epsilon} \log d\Pb_\epsilon \Bigm|_{\epsilon=0} \right) \ d\Pb(z)
\end{equation}
for any regular parametric submodel $\{\Pb_\epsilon : \epsilon \in \R\}$ with scores in the tangent space. The \textit{efficient influence curve} is the unique influence curve that is also an element of the tangent space (and thus can be defined as the projection of any influence curve on the tangent space $\Pi(\varphi \mid \mathcal{T})$). It is also the curve with the smallest covariance $P(\varphi_\eff^2) \leq P(\varphi^2)$ for all $\varphi$, and can further be expressed as $\varphi_\eff=P(s_\eff^2)^{-1} s_\eff$, where $s_\eff$ is the {efficient score}, i.e., the projection of the score onto the tangent space, i.e., $s_\eff=\Pi(s_\psi \mid \mathcal{T}_\eta^\perp) = s_\psi-\Pi(s_\psi \mid \mathcal{T}_\eta)$. 

\section{Finding \& Using Influence Curves}

Characterizing the influence curves, i.e., the putative influence functions, for a particular functional $\psi(P)$ and model $\mathcal{P}$ is a critical task with very important ramifications. The efficient influence curve gives the efficiency bound for estimating $\psi$, thus providing a benchmark against which estimators can be compared. Perhaps more importantly, influence curves can be used to construct estimators with very favorable properties, such as double robustness and general second-order bias, and which improve on naive plug-in estimators that require stronger smoothness conditions and often impractical bandwidth choices. In particular, given an influence curve $\varphi(Z;\eta,\psi)$ depending on nuisance functions $\eta$ and the parameter of interest $\psi$, one can construct an estimator $\hat\psi$ by solving the estimating equation $\Pn\{ \varphi(Z;\hat\eta,\psi)\}=0$ based on the estimated influence curve. The resulting estimator $\hat\psi$ can be shown to have influence function $\varphi$ using the logic from Section 3. 

In semiparametric models the tangent space is a proper subspace of $L_2(P)$, and deriving influence curves can be a delicate task that generally requires characterizing the nuisance tangent space and its complement. In nonparametric models the situation is often more hopeful: then there is only one influence curve, it is efficient, and it can often be computed directly via derivative calculations. For example, one can temporarily assume discrete data and compute the Gateaux derivative $\frac{d}{d\epsilon} \psi(d\Pb_\epsilon)|_{\epsilon=0}$ along the submodel $d\Pb_\epsilon(z) = (1-\epsilon) d\Pb(z) + \epsilon \delta_{z'}$ for $\delta_z$ the Dirac measure at $Z=z$, which yields the influence curve evaluated at $z'$.  Then it is typically straightforward to see the corresponding influence curve in the general case, e.g., by replacing probability mass functions with densities. For the treatment effect example, these calculations show that the efficient influence curve for the effect $\psi$ is exactly the influence function $\varphi_{dr}^*-\psi$ given in Section 3.

\stepcounter{section}
\printbibliography[title={\thesection \ \ \ References}]

\end{document}